\documentclass[aps, prb, preprint, amssymb, amsmath, showpacs, superscriptaddress]{revtex4}

\usepackage{graphicx}
\usepackage{dcolumn}
\usepackage{bm}

\begin{document}

\title{Half-Heusler Topological Insulators: A First-Principle Study with the Tran-Blaha Modified Becke-Johnson Density Functional}

\author{Wanxiang Feng}
\affiliation {Institute of Physics, Chinese Academy of Sciences, Beijing 100190, China}

\author{Di Xiao}
\affiliation {Materials Science \& Technology Division, Oak Ridge National Laboratory, Oak Ridge, TN 37831, USA}

\author{Ying Zhang}
\affiliation {Department of Physics, Beijing Normal University, Beijing 100875, China}
\affiliation {Department of Physics, The University of Texas at Austin, Austin, TX 78712, USA}

\author{Yugui Yao}
\affiliation {Institute of Physics, Chinese Academy of Sciences, Beijing 100190, China}
\affiliation {Department of Physics, The University of Texas at Austin, Austin, TX 78712, USA}

\date{\today}

\begin{abstract}
We systematically investigate the topological band structures of half-Heusler compounds using first-principles calculations. The modified Becke-Johnson exchange potential together with local density approximation for the correlation potential (MBJLDA) has been used here to obtain accurate band inversion strength and band order. Our results show that a large number of half-Heusler compounds are candidates for three-dimensional topological insulators.  The difference between band structures obtained using the local density approximation (LDA) and MBJLDA potential is also discussed.
\end{abstract}


\maketitle

\section{INTRODUCTION}

Motivated by their potential applications in spintronics and quantum computing~\cite{Moore2010}, the search for three-dimensional topological insulators (3DTI) has attracted considerable theoretical and experimental interest~\cite{Fu2007a,Hsieh2008,Zhang2009,Xia2009,Chen2009,Yan2010,Lin2010a,Sato2010,Chen2010, Xiao2010,Lin2010b,Chadov2010}.  These materials are called ``topological'' because they are distinguished from ordinary insulators by the so-called $Z_2$ topological invariants associated with the bulk band structure~\cite{Kane2005,Fu2006}.  On the theory side, the state-of-the-art first-principles calculations guided by topological band theory~\cite{Fu2007b,Moore2007,Roy2009} has provided a powerful tool for uncovering new families of 3DTI.  Based on this approach, it has been recently predicted that ternary half-Heusler compounds can realize the topological insulating phase with proper strain engineering~\cite{Xiao2010,Lin2010b,Chadov2010}.

There are several ways to determine the band topology of an insulator.  Intuitively, one can count the number of band inversions within the entire Brillouin zone -- an \emph{odd} number indicates that the material may be a 3DTI~\cite{Fu2007a,Lin2010a}.  This method depends on an accurate interpretation of the atomic origin of the bands and is better suited for crystals with a high-symmetry lattice.  A more rigorous method is to directly evaluate the $Z_2$ topological invariants.  For materials with inversion symmetry, the parity criteria developed by Fu and Kane~\cite{Fu2007a} can be readily applied.  On the other hand, if the inversion symmetry is absent, one must resort to the more elaborated lattice computation of the $Z_2$ invariants~\cite{Fukui2007}.  We have recently applied this method to study the distorted half-Heuslers~\cite{Xiao2010} and the noncentrosymmetric chalcopyrite compounds~\cite{Feng2010}.  In practical calculations, both methods require an accurate knowledge of the bulk band structures.  So far, previous works on 3DTI have employed either the local density approximation (LDA)~\cite{Kohn1965,Perdew1992} or generalized gradient approximation (GGA)~\cite{Perdew1996} for the exchange-correlation potential. However, it is well known that these approximations have the tendency to underestimate the bandgap.  In particular, if the material has a small positive bandgap, this underestimation may yield a negative value.  Therefore, these approximations may falsely predict an inverted band structure when the band order is actually normal.

Recently, a new semilocal potential that combines the modified Becke-Johnson (MBJ) exchange potential and the LDA correlation potential, called MBJLDA~\cite{Tran2009}, was proposed to obtain accurate bandgaps and band order. The MBJLDA potential are computationally as cheap as LDA or GGA, but it has similar precision compared with the more expensive hybrid functionals and GW method.  It has been demonstrated~\cite{Tran2009} that the MBJLDA potential can be used to describe many types of solids, including wide bandgap insulators, small bandgap $sp$ semiconductors, and strongly correlated $3d$ transition-metal oxides. More importantly, the MBJLDA potential can effectively mimic the behavior of orbital-dependent potential around the bandgap, so it is expected to obtain accurate positions of states near the band-edge, which are the keys to determine the band inversion and the band topology.

In this work we perform a systematic investigation of the band topology of the half-Heusler family using the MBJLDA potential, and compare it with the LDA result.  The improved accuracy of the MBJLDA potential over LDA is tested in the ordinary insulator CdTe and the topologically nontrivial 3D-HgTe,  whose band structures are known experimentally and can be used as a benchmark.  We then focus on the ternary half-Heusler compounds.  By calculating the band inversion strength, we confirm our previous prediction that a large number of half-Heuslers are possible candidates for 3DTI or 3D topological metal~\cite{Xiao2010}.  The difference between LDA and the MBJLDA potential is also discussed and a clear discrepancy is found between these two methods.  The sensitivity of the calculated band structure due to the type of exchange-correlation potential calls for more detailed experimental works.

\begin{figure}
  \includegraphics[width=3.5in]{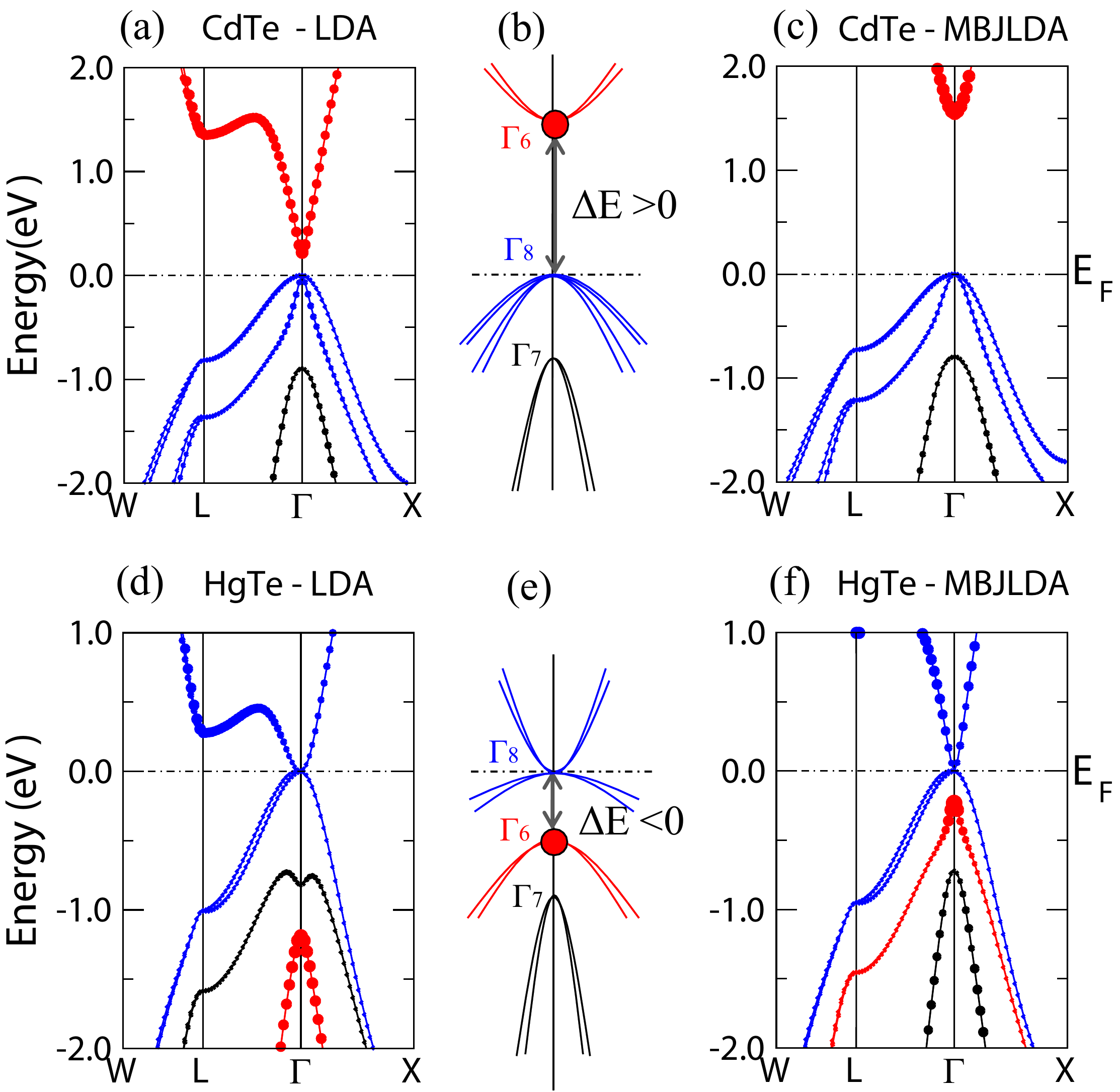}
  \caption{(color online)Band structure of CdTe and HgTe. (a) LDA and (c) MBJLDA results of CdTe; (d) LDA and (f) MBJLDA results of HgTe. The $\Gamma_6$, $\Gamma_7$, and $\Gamma_8$ state are denoted by red, black, and blue color, respectively. The size of dots is proportional to the probability of s-orbit projection. (b) and (e) is schematically experimental band structure of CdTe and HgTe, the red solid dots denote the s-orbit originated $\Gamma_6$ state. The band inversion strength is defined as $\Delta E = E_{\Gamma_6} - E_{\Gamma_8}$. The calculated and experimental energy of $\Gamma_6$, $\Gamma_7$, and $\Gamma_8$ state are listed in Table ~\ref{tab:energy}.}\label{fig:order}
\end{figure}

\section{Method}

The band structure calculations in this work were performed using full-potential linearized augmented plane-wave (FP-LAPW) method~\cite{Singh1994}, implemented in the package \textsc{wien2k}~\cite{Blaha2001}. A converged ground state was obtained using 10,000 k points in the first Brillouin zone with $K_{max}R_{MT}=9.0$, where $R_{MT}$ represents the muffin-tin radius and $K_{max}$ is the maximum size of reciprocal-lattice vectors. Wave functions and potentials inside the atomic sphere are expanded in spherical harmonics up to $l=10$ and $4$, respectively. Spin-orbit coupling is included by a second-variational procedure~\cite{Singh1994}, where states up to 9 Ry above fermi level are included in the basis expansion, and the relativistic $p_{1/2}$ corrections were also considered for $5p$ and $6p$ orbit in order to improve the accuracy~\cite{Larson2003}.

\section{Results and Discussions}

\subsection{CdTe and HgTe}

We first test the MBJLDA potential for the binary compounds CdTe and HgTe. The energy bands of both compounds at the $\Gamma$ point near the Fermi level split into $\Gamma_6$ (2-fold degenerate), $\Gamma_7$ (2-fold degenerate), and $\Gamma_8$ (4-fold degenerate) states due to the zinc-blende crystal symmetry and strong spin-orbit interaction.  The experimental band order of CdTe and HgTe are (from high to low energy) $\Gamma_6$, $\Gamma_8$, $\Gamma_7$ (Ref.~\onlinecite{Lemasson1982,Niles1991}) and $\Gamma_8$, $\Gamma_6$, $\Gamma_7$ (Ref.~\onlinecite{Orlowski2000}), respectively.  They are schematically shown in Fig.~\ref{fig:order}(b) and \ref{fig:order}(e).  From the viewpoint of band topology, the main difference between these two compounds is that CdTe possesses a normal band order, i.e., the $s$-like $\Gamma_6$ state sits above the $p$-like $\Gamma_8$ state, while HgTe possesses an inverted band order in which the $\Gamma_6$ state is occupied and sits below the $\Gamma_8$ state.  We then define the band inversion strength $\Delta E$ as the energy differences between these two states, i.e.
\begin{equation} \label{band}
\Delta E = E_{\Gamma_6} - E_{\Gamma_8} \;.
\end{equation}
A negative $\Delta E$ typically indicates that the materials are in a topologically nontrival phase, while those  with a positive $\Delta E$ are in a topologically trivial phase.

Figure~\ref{fig:order} shows the LDA and MBJLDA band structures of CdTe and HgTe at their experimental lattice constants 6.48\AA and 6.46\AA, respectively~\cite{Woolley1960}. The calculated energy of the $\Gamma_6$, $\Gamma_7$, and $\Gamma_8$ states together with their experimental values are listed in Table~\ref{tab:energy}.    For CdTe, the LDA potential yields a serious underestimation of $\Delta E$.  For HgTe, it obtains the wrong band order of the $\Gamma_6$ and $\Gamma_7$ states (but it does predict an inverted band order).  On the other hand, the MBJLDA result shows an excellent agreement with experiments for both compounds.  We therefore conclude that the MBDLDA potential is better suited to calculate the topological band structure.

\begin{table}
\caption{The energy of $\Gamma_6$, $\Gamma_7$, and $\Gamma_8$ state of CdTe and HgTe calculated with LDA and MBJLDA at their experimental lattice constants 6.48\AA and 6.46\AA, respectively~\cite{Woolley1960}. Experimental values are also shown for comparison. The fermi level are always located on $\Gamma_8$ state, and $E_{\Gamma_8}$ are set to zero. The energy units are eV.}\label{tab:energy}
\begin{ruledtabular}
\begin{tabular}{ccddd}
\multicolumn{1}{c}{\textrm{}} &
\multicolumn{1}{c}{\textrm{}} &
\multicolumn{1}{c}{\textrm{LDA}} &
\multicolumn{1}{c}{\textrm{MBJLDA}} &
\multicolumn{1}{c}{\textrm{Experiment}} \\
\hline
     & $E_{\Gamma6}$ &  0.221 &  1.549 &  1.475\footnotemark[1] \\
CdTe & $E_{\Gamma8}$ &  0.0   &  0.0   &  0.0                   \\
     & $E_{\Gamma7}$ & -0.897 & -0.795 & -0.95\footnotemark[2]  \\
\hline
     & $E_{\Gamma8}$ &  0.0   &  0.0   &  0.0                   \\
HgTe & $E_{\Gamma6}$ & -1.191 & -0.234 & -0.29 \footnotemark[3] \\
     & $E_{\Gamma7}$ & -0.822 & -0.721 & -0.91 \footnotemark[3] \\
\end{tabular}
\end{ruledtabular}
\footnotetext[1]{Reference~\onlinecite{Lemasson1982}.}
\footnotetext[2]{Reference~\onlinecite{Niles1991}.}
\footnotetext[3]{Reference~\onlinecite{Orlowski2000}.}
\end{table}

\subsection{Half-Heuslers}

\begin{figure}
\includegraphics[width=6cm]{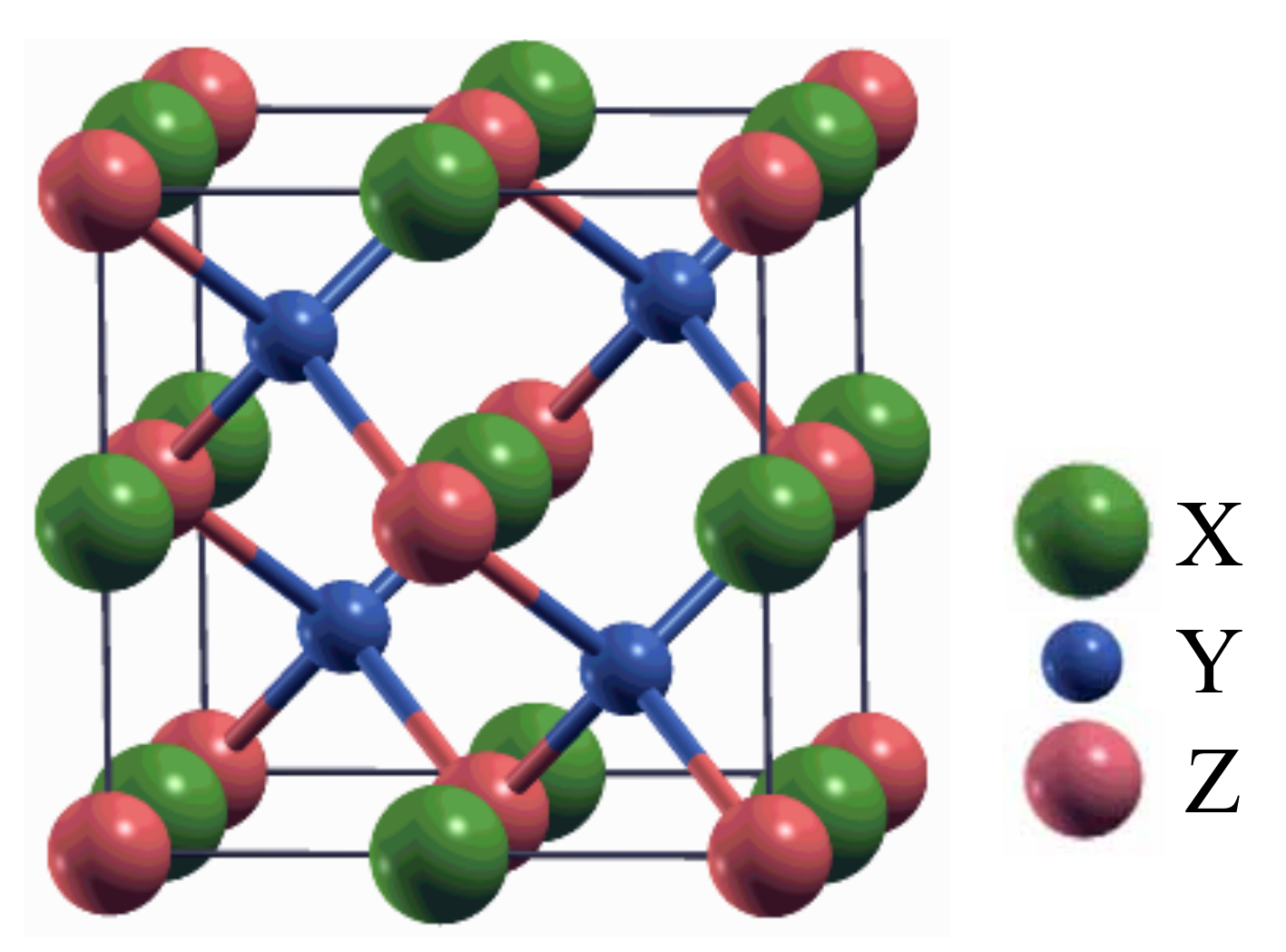}
\caption{\label{fig:struct}(color online) Crystal structure of
  half-Heusler compound XYZ in the $F\bar{4}3m$ space group. Green
  spheres at (0.5,0.5,0.5) are atom X, dark blue spheres at
  (0.25,0.25,0.25) are atom Y, and pink spheres at (0,0,0) are atom
  Z.}
\end{figure}

\begin{figure*}
  \includegraphics[width=7.0in]{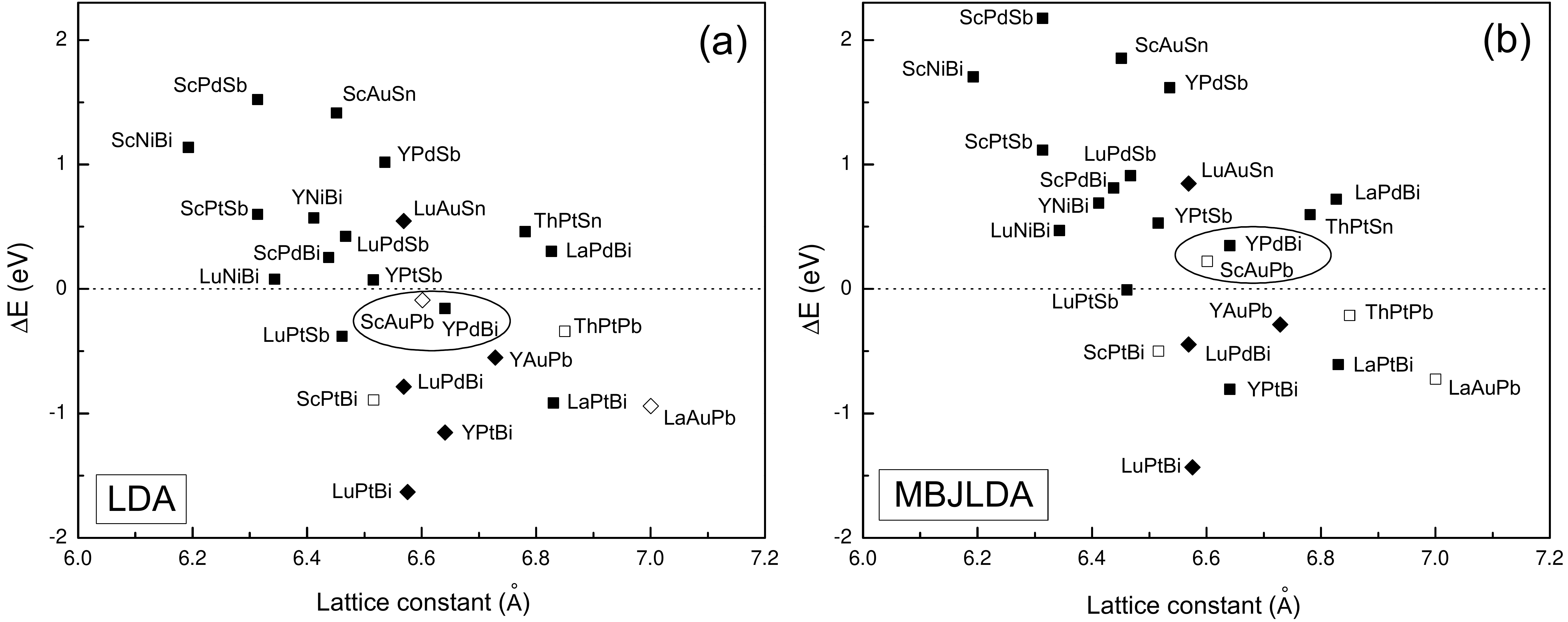}
  \caption{Band inversion strength of half-Heusler compounds calculated by LDA (a) and MBJLDA (b). The materials with open symbols indicate that there is not experimental reports, and the lattice constant are obtained by first-principles total energy minimization. The topological insulator or metal (after applying a proper uniaxial strain) with $\Delta E < 0$ are located below the horizontal line; and the ordinary insulator or metal with $\Delta E > 0$ are located above the horizontal line. The squares mark insulator or zero-bandgap semiconductor and diamonds mark the metal. The materials enclosed by circle indicate that their topological phase predicted change by using different exchange-correlation potential.}\label{fig:map}
\end{figure*}

Having established the improved accuracy of the MBJLDA potential over LDA, we now turn to the ternary half-Heusler compounds described by space group $F\bar{4}3m$. The chemical formula of these materials is $XYZ$, where $X$ and $Y$ are transition or rare earth metals and $Z$ a heavy element. It can be regarded as a hybrid compound of $XZ$ with rock-salt structure, and $XY$ and $YZ$ with the zinc-blende structure  (see Fig.~\ref{fig:struct}). In our band structure calculations, the lattice constants are taken from experimental data library~\cite{Villars1991,springer}.  For compounds without experimental lattice constant, we use the value obtained by total energy minimization in first-principles calculations.

\begin{figure*}
  \includegraphics[width=7.0in]{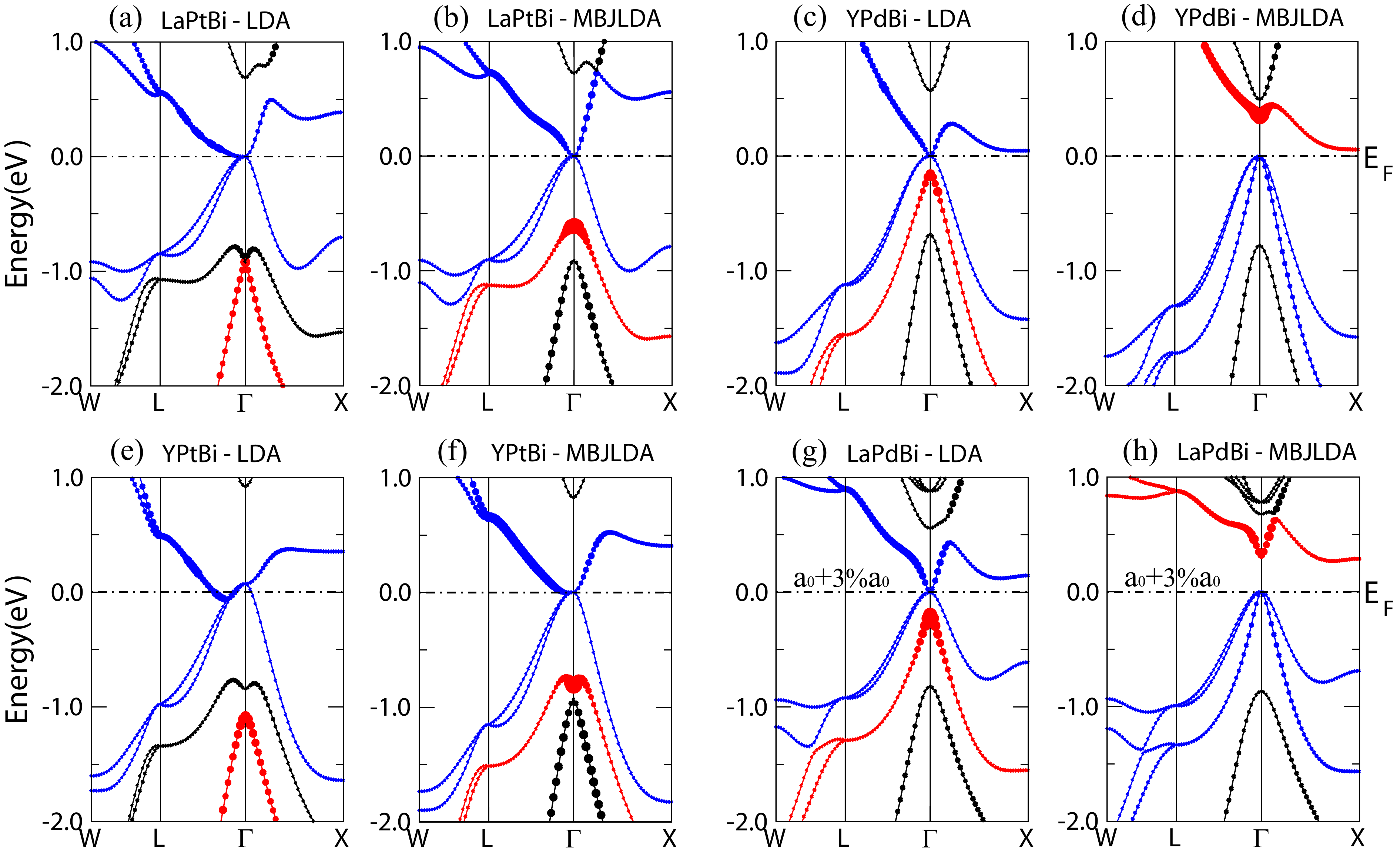}
  \caption{(color online) LDA and MBJLDA band structure of half-Heusler compounds LaPtBi(a)(b), YPdBi(c)(d), YPtBi(e)(f), and LaPdBi(g)(h). The labels are the same as Fig. ~\ref{fig:order}.}\label{fig:bands}
\end{figure*}

The band structures of half-Heuslers are very similar to CdTe/HgTe.  In particular, the low-energy electron dynamics is dominated by energy bands at the $\Gamma$ point.  Therefore, the band topology of half-Heuslers can be characterized by the band inversion strength $\Delta E$ defined in Eq.~\eqref{band}.  In order to  systemically explore their topological phase, we have calculated $\Delta E$ by both LDA and MBJLDA for 24 half-Heusler compounds, as shown in Fig. ~\ref{fig:map}(a) and (b), respectively.  Following our previous work (Ref.~\onlinecite{Xiao2010}), we have also calculated the $Z_2$ invariant for all half-Heusler compounds investigated here.  The calculated $Z_2$ invariant agrees with the intuitive band inversion picture, that is, materials with negative $\Delta E$ are in topologically nontrivial phase.  As shown in Fig.~\ref{fig:map}(b), we have identified 9 half-Heusler compounds, LuPtSb, LaPtBi, YPtBi, ScPtBi, ThPtPb, LaAuPb, YAuPb, LuPdBi, LuPtBi  as possible candidates for topological insulators or topological metal. (Note that   ScPtBi, ThPtPb, LaAuPb here are virtual compounds.)  Here, the band structures of the first 6 compounds are with zero band-gap similar to that of HgTe, and the last 3 compounds possess the nontrivial metallic  band structures similar to Fig.~\ref{fig:bands}(e)  or the conduction bands  droping below the Fermi level near the X point.

As shown in section III.A, the LDA result generally yields a smaller value of the band inversion strength $\Delta E$ when compared with the MBJLDA result, which agrees much better with the experimental data.  We emphasize that, unlike the usual energy gaps, here $\Delta E$ can take on both positive and negative values, hence a positive $\Delta E$ is always larger than a negative one.  This trend is also confirmed for half-Heuslers.  As shown in Fig.~\ref{fig:map}, $\Delta E$ by MBJLDA is always larger than that by LDA.  In particular, if LDA calculation predicts a small negative $\Delta E$, it may become positive when using MBJLDA potential, such as ScAuPb, and YPdBi.  This suggests that when predicting topological phases, one must be careful with the types of exchange-correlation potentials. 

To illustrate the general difference between the LDA and MBJLDA results, we present four scenarios shown in Fig.~\ref{fig:bands} using some typical examples.  Both LDA and MBJLDA calculations predict that LaPtBi [Fig.~\ref{fig:bands}(a) and ~\ref{fig:bands}(b)] has an inverted band structure, which is very similar to HgTe. The main difference is the band order exchange between $\Gamma_6$ and $\Gamma_7$ state. Although the band topology will not change, the LDA result clearly underestimates the $\Delta E$ due to the downward movement of $\Gamma_6$ state. For YPdBi, the problem is much more severe as the two potentials give different band topology.  As shown in Fig.~\ref{fig:bands}(c) and \ref{fig:bands}(d), the LDA result shows an inverted band order but the MBJLDA result shows a normal band order.  The two types of exchange-correlation potentials also affects the electronic behavior around the Fermi level as shown for YPtBi in Fig.~\ref{fig:bands}(e) and \ref{fig:bands}(f).  The metallic phase predicted by LDA calculation becomes a zero-bandgap semiconductor in MBJLDA calculation.  The band order of $\Gamma_6$ and $\Gamma_7$ is also exchanged, just like LaPtBi. 

 \begin{figure}
  \includegraphics[width=8.0cm]{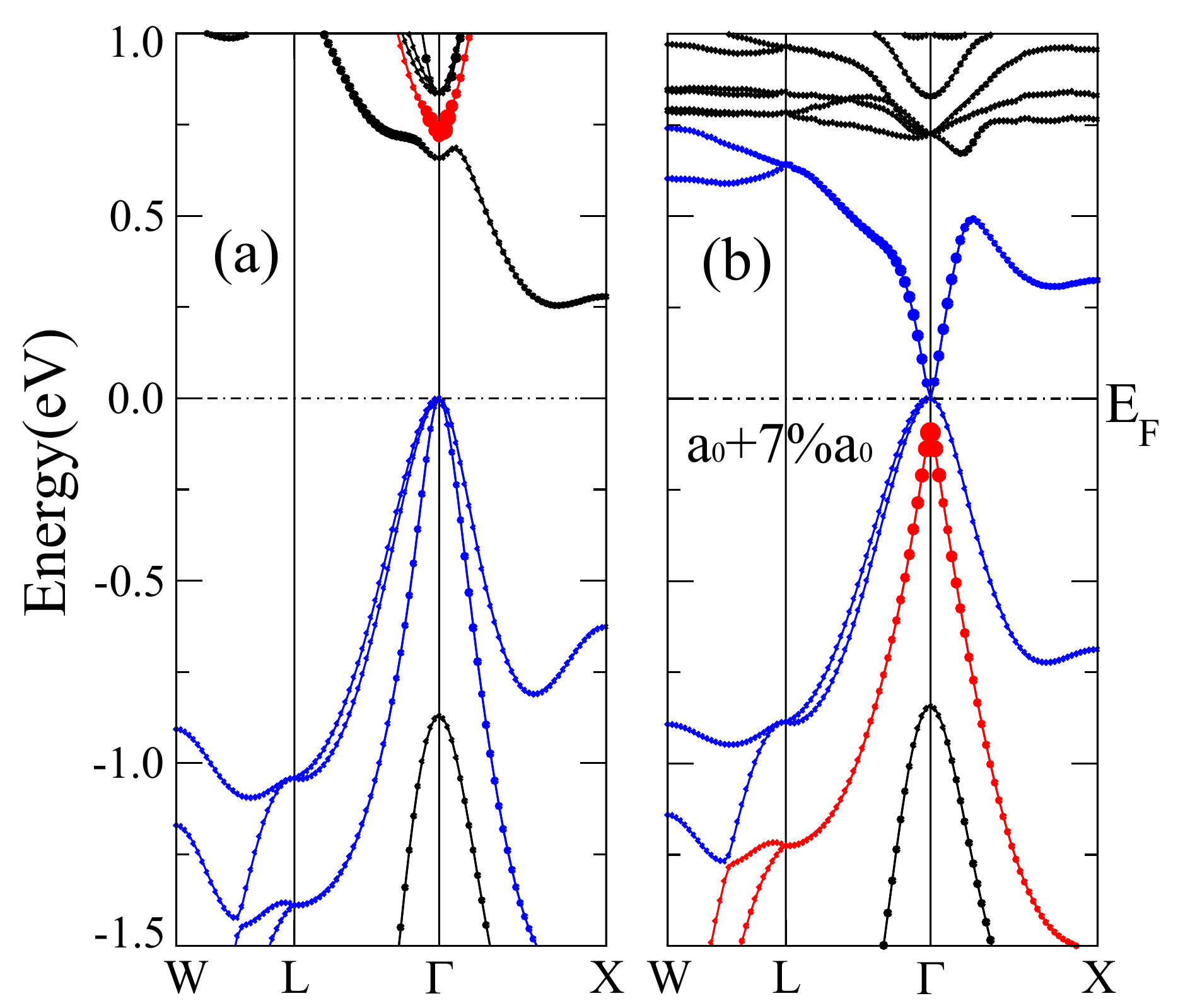}
  \caption{(color online) Band structures of LaPdBi by MBJLDA (a) without  and (b) with a 7\% hydrostatic strain. The application of a hydrostatic strain causes the s-like
$\Gamma_6$ bands (marked by red dots) to jump below the valence band top, providing the necessary band inversion that leads to a nontrivial topological
order. The other labels are the same as Fig. ~\ref{fig:order}.}\label{fig:LaPdBi}
\end{figure}

As mentioned in our previous work~\cite{Xiao2010},  for half-Heusler compounds with small band gaps, the nontrivial topologically phase can be generally realized by applying hydrostatic strain to change the band order.  
 For example,  LaPdBi has normal band order in its native state, a $7\%$ change in the lattice constant converts the trivial topological phase into a nontrivial topological phase (Fig.~\ref{fig:LaPdBi}). Again, there are different effects on the band order under the same hydrostatic strain when using different exchange-correlation potentials.  For example, when one stretches lattice constant to $a=a_0+3\%a_0$,  LDA calculation will yield an inverted band structure [Fig. ~\ref{fig:bands}(g)], while it still has a normal band order by MBJLDA calculation[Fig. ~\ref{fig:bands}(h)].

\section{Summary}

In summary, we have systematically investigated the topological band structure of the half-Heusler family using both LDA and MBJLDA exchange-correlations potential. Our result shows that a large number of half-Heusler compounds are candidates for three-dimensional topological insulators.  We also discussed the main differences between the LDA and MBJLDA potentials

\begin{acknowledgments}
This work was supported by NSF of China (10674163, 10974231) and the MOST Project of China (2007CB925000), Welch Foundation (F-1255) and DOE (DE-FG02-02ER45958, Division of Materials Science and Engineering), and by Supercomputing Center of Chinese Academy of Sciences(SCCAS) and Texas Advanced Computing Center(TACC).

\end{acknowledgments}

\end{document}